\documentclass[superscriptaddress,prc,preprint,showpacs]{revtex4}

\date{May 8, 2003}
\usepackage{hyperref}
\usepackage{graphicx}
\usepackage{amssymb}
\usepackage{amsmath}

\newcommand{\calcium}[1][40]{{}^{#1}\mathrm{Ca}}

\newcommand{\MSU}{National Superconducting Cyclotron Laboratory and
  Department of Physics and Astronomy, Michigan State University, East
  Lansing, Michigan 48824}
\newcommand{\WIS}{
  Department of Physics, University of Wisconsin,
  Madison, WI 53706, USA}
\newcommand{\Tohoku}{Department of Physics, Tohoku University, Sendai
  980-8578, Japan}

\begin{document}

\title{Isospin fractionation and isoscaling in dynamical
nuclear collisions}

\author{Akira Ono}
\affiliation{\Tohoku}
\author{P. Danielewicz}
\affiliation{\MSU}
\author{W. A. Friedman}
\affiliation{\WIS}
\author{W. G. Lynch}
\affiliation{\MSU}
\author{M. B. Tsang}
\affiliation{\MSU}

\begin{abstract}
Isoscaling is found to hold for fragment yields in the antisymmetrized
molecular dynamics (AMD) simulations for collisions of calcium
isotopes at 35 MeV/nucleon. This suggests the applicability of
statistical considerations to the dynamical fragment emission. The
observed linear relationship between the isoscaling parameters and the
isospin asymmetry of fragments supports the above suggestion. The
slope of this linear function yields information about the symmetry
energy in low density region where multifragmentation occurs.
\end{abstract}

\pacs{25.70.Pq}
\maketitle


In typical intermediate energy nuclear collisions, numerous fragments
of intermediate size are produced in addition to light particles
\cite{Multifrag}.  The multifragmentation phenomenon is believed to be
related to the liquid-gas phase-coexistence in low density expanding
nuclear matter.  In a two-component system with more neutrons than
protons ($N^\text{tot}>Z^\text{tot}$) in equilibrium, the gas phase
becomes more neutron-rich than the liquid phase \cite{MULLER}.  This
fractionation phenomenon should reflect the features of the symmetry
energy in nuclear matter.

Recently, a scaling relation
\begin{equation}
Y_2(N,Z)/Y_1(N,Z)\propto e^{\alpha N+\beta Z}\label{eq:isoscaling}
\end{equation}
has been observed \cite{XU-isoscale} in the measured fragment yields
$Y_i(N,Z)$ for two similar systems $i=1,2$ with different neutron to
proton ratios.  This phenomenon is called isoscaling.  If one assumes
thermal and chemical equilibrium, the isoscaling parameters $\alpha$
and $\beta$ are related to the neutron-proton content of the emitting
source.  In fact, statistical models have successfully explained the
isoscaling data \cite{TSANG-statmodel}.  However, as fragments are
formed during a dynamical evolution of the collision system,
multifragmentation should be understood in the dynamical models as
well.  In fact, a stochastic mean field model has predicted very large
scaling parameters for the dynamically produced fragments in the model
\cite{SMF-isoscale}.  Such result is difficult to understand without
any dynamical effects. It is also important to determine if the
scaling parameters in the data can be directly related to the
asymmetry term of the equation of state (EOS) of nuclear matter in
equilibrium, for a dynamic production.

To explore whether any kind of equilibrium is achieved regarding the
isospin fractionation and the fragmentation in dynamical nuclear
collisions, we compare the result of the antisymmetrized molecular
dynamics (AMD) simulation with what is expected under a statistical
assumption.  We first derive, under an equilibrium assumption, a
linear relation between the isoscaling parameter $\alpha$ and the
fragment isospin asymmetry $(Z/A)^2$. We then test such a relationship
using results from the AMD simulations.  By studying the dependence on
the asymmetry term of the effective force, we will explore whether the
relation is useful for assessing the asymmetry term of the EOS.


AMD is a microscopic model for following the time evolution of nuclear
collisions \cite{ONOab,ONOh,ONOi}.  It represents the colliding system
in terms of a fully antisymmetrized product of Gaussian wave packets.
Through the time evolution, the wave packet centroids move according
to an equation of motion.  Besides, the followed state of the
simulation branches stochastically and successively into a huge number
of reaction channels.  The branching is caused by the two-nucleon
collisions and by the splittings of the wave packet.  The interactions
are parametrized in the AMD model in terms of the effective force
between nucleons and the two-nucleon collision cross sections.

\begin{figure}
\includegraphics[width=0.48\textwidth]{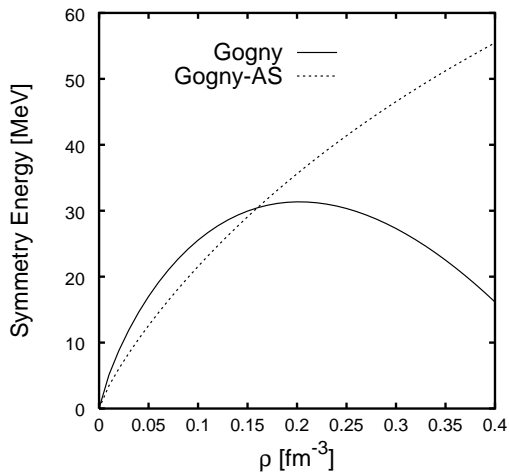}
\caption{\label{fig:symeng} Density dependence of the symmetry energy
of nuclear matter for the Gogny force (solid line) and for the
Gogny-AS force (dashed line).}
\end{figure}

We perform reaction simulations employing two different effective
forces in order to study effects of the asymmetry term.  One is the
usual Gogny force \cite{GOGNY}, consistent with the saturation of
symmetric nuclear matter at the incompressibility $K=228$ MeV.  The
force is composed of finite-range two-body terms and of a
density-dependent term of the form
$t_3\rho^{1/3}(1+P_\sigma)\delta(\mathbf{r}_1-\mathbf{r}_2)$, where
$P_\sigma$ is the spin exchange operator and $t_3$ is a coefficient.
The second force (called Gogny-AS force) is obtained by modifying the
Gogny force with
\begin{equation}
V_\text{Gogny-AS}=V_\text{Gogny}
-(1-x)t_3\Bigl(\rho(\mathbf{r}_1)^{1/3}-\rho_0^{1/3}\Bigr)
P_\sigma\delta(\mathbf{r}_1-\mathbf{r}_2),
\end{equation}
where $x=-\frac{1}{2}$ and $\rho_0=0.16\ \text{fm}^{-3}$.  The two
forces coincide at $\rho=\rho_0$.  Furthermore, they produce the same
EOS of symmetric nuclear matter at all density.  However, the two
forces produce different density dependences of the symmetry energy,
as shown in Fig.\ \ref{fig:symeng}.  The choice of $x=-\frac{1}{2}$
has been made to ensure that the part of the symmetry energy from the
direct term is proportional to the density \cite{DITORO-force}. At
densities below $\rho_0$, the Gogny force has somewhat higher symmetry
energy than the Gogny-AS force. At densities above $\rho_0$, the
Gogny-AS symmetry energy continues to rise while the Gogny symmetry
energy begins to fall, so that significant differences develop.

The AMD simulations were performed for $\calcium[40]+\calcium[40]$,
$\calcium[48]+\calcium[48]$ and $\calcium[60]+\calcium[60]$ collisions
at the incident energy $E/A=35$ MeV/nucleon and zero impact parameter.
The version of AMD of Ref.\ \cite{ONOi} was utilized.  It has been
demonstrated that an equivalent version of AMD, for the present
purposes, reproduces the experimental data of various fragment
observables in $\calcium[40]+\calcium[40]$ at the same energy of 35
MeV/nucleon with the Gogny force \cite{ONOh,WADAa}.  Each studied
collision event was started by boosting two nuclei with centers
separated by 9 fm.  The dynamical simulation was continued until
$t=300$ fm/$c$.  About 1000 events were generated for each system.

In central collisions, as shown in a previous paper \cite{ONOh,WADAa},
two nuclei basically penetrate each other and many fragments are
formed not only from the projectile-like and target-like parts but
also from within the neck region between the two residues.  The
nuclear matter seems to be strongly expanding, one-dimensionally, in
the beam direction.

\begin{figure}
\includegraphics[width=0.48\textwidth]{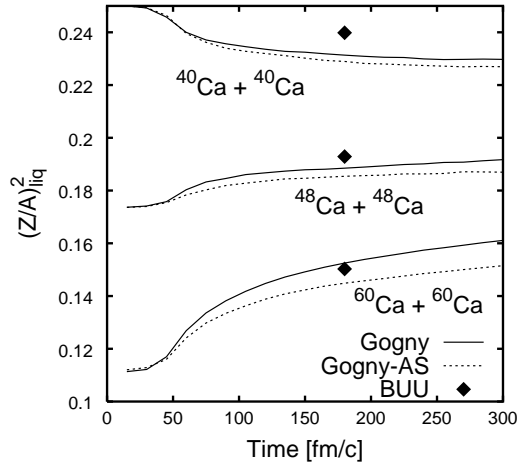}
\caption{\label{fig:reszoasq2} $(Z/A)^2$ of the liquid part of the
system as a function of time for the three reactions systems.  The AMD
results are represented by the solid and dashed lines, respectively,
for the Gogny and Gogny-AS forces.  Late-time BUU results are
represented by filled diamonds.}
\end{figure}

\begin{figure}
\includegraphics[width=0.48\textwidth]{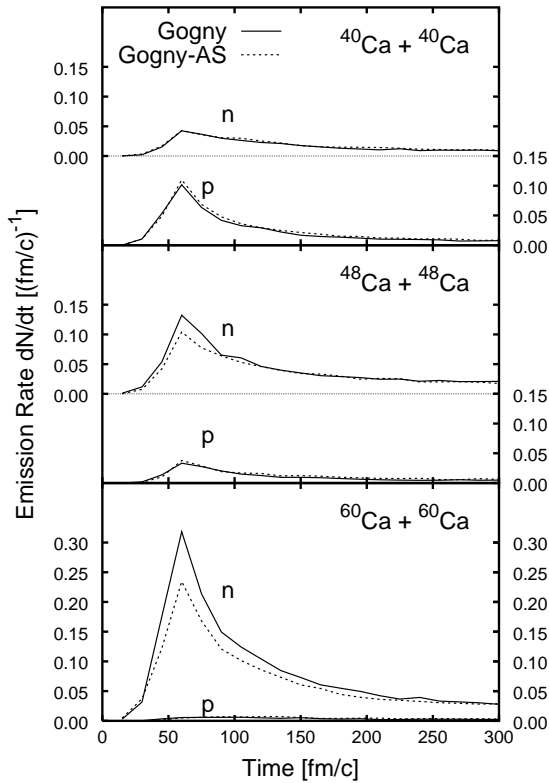}
\caption{\label{fig:amdemit} Neutron and proton emission rates
described by the left- and right-hand scales, respectively, for the
three reaction systems as a function of time.  The results of AMD
simulations with the Gogny and Gogny-AS forces are, respectively,
represented by the solid and dashed lines.  }
\end{figure}

For the intermediate states, we define the liquid part as the part of
the system to be composed of the fragments with $A>4$ and any two wave
packets whose spatial separation is less than 3 fm are treated as
belonging to the same fragment.  In the context of the results of
Ref.\ \cite{TSANG-statmodel}, Fig.\ \ref{fig:reszoasq2} shows the time
evolution of the isospin asymmetry $(Z/A)^2$ of the liquid part for
the three reaction systems.  At the initial value ($t\sim0$),
$(Z/A)^2_\text{liq}$ is $(Z/A)^2$ of the initial nuclei.  For the
neutron-rich systems, $(Z/A)^2_\text{liq}$ increases rapidly before
$t\sim 100$ fm/$c$, and then it continues to increase gradually.  This
effect can be regarded as the isospin fractionation because the liquid
part is getting less neutron-rich and the gas part is getting more
neutron-rich.  Similar fractionation effects are found in other
dynamical model simulations \cite{DITORO-force,BAOANLI}. The diamond
points in Fig.\ \ref{fig:reszoasq2} are the results obtained from the
Boltzmann-Uehling-Uhlenbeck calculations \cite{BUU-pawel} with an
interaction symmetry energy of $12.125(\rho/\rho_0)$ MeV.  The isospin
fractionation has a clear dependence on the asymmetry term of the
effective force.  The Gogny force (solid lines) always yields a system
with larger $(Z/A)^2_\text{liq}$ than the Gogny-AS force (dashed
lines).  The complementary effect of fractionation can also be found
in the gas-phase information, such as the neutron and proton emission
rates in Fig.\ \ref{fig:amdemit}.  While for the symmetric system
($\calcium[40]+\calcium[40]$) more protons are emitted than neutrons,
because of the Coulomb force, for the very neutron-rich system
($\calcium[60]+\calcium[60]$) much more neutrons are emitted.

\begin{figure}
\begin{center}
\includegraphics[width=0.47\textwidth]{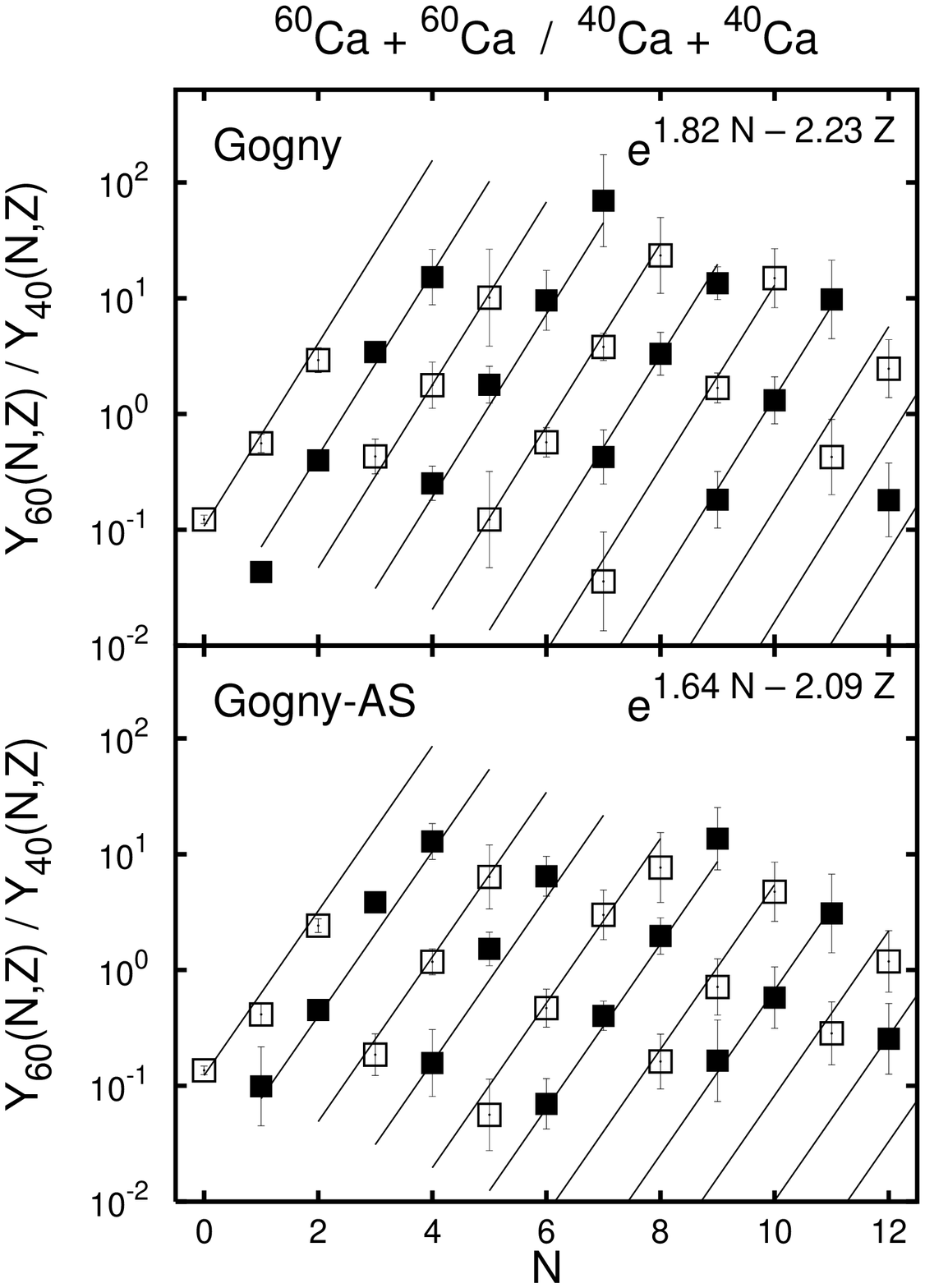}
\end{center}
\caption{\label{fig:isoscale2-6040} The fragment yield ratio between
  the AMD simulations of central $\calcium[60]+\calcium[60]$ and
  $\calcium[40]+\calcium[40]$ collisions at 35 MeV/nucleon, at time
  $t=300$ fm/$c$.  The top and bottom panels show, respectively, the
  results obtained using the Gogny and Gogny-AS forces.  The extracted
  isoscaling parameters are $\alpha=1.82\pm0.06$ and
  $\beta=-2.23\pm0.08$ for the Gogny force, and $\alpha=1.64\pm0.05$ and
  $\beta=-2.09\pm0.07$ for the Gogny-AS force.  }
\end{figure}

\begin{figure}
\begin{center}
\includegraphics[width=0.47\textwidth]{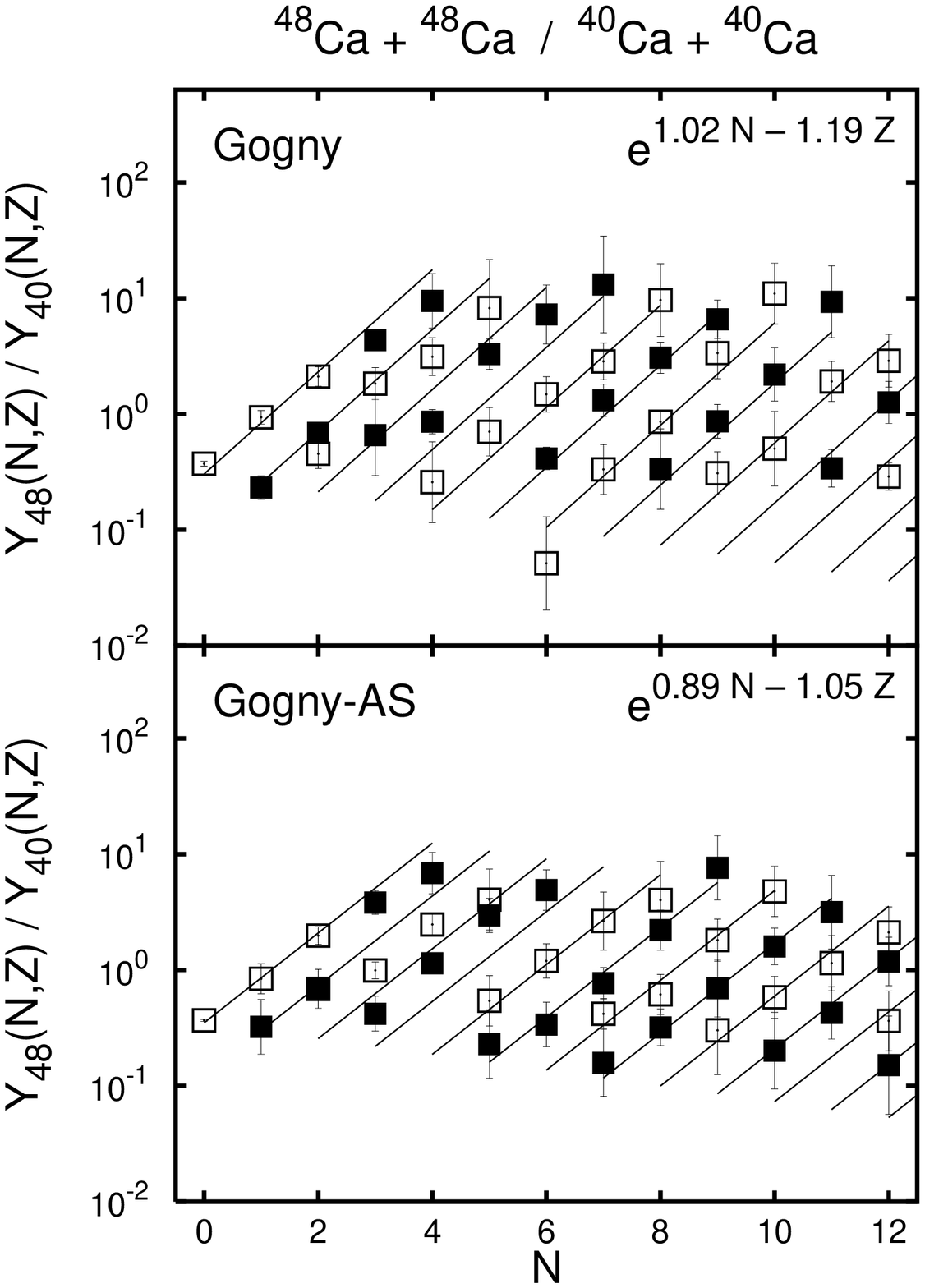}
\end{center}
\caption{\label{fig:isoscale2-4840} The same as Fig.\
  \ref{fig:isoscale2-6040} but for the $\calcium[48]+\calcium[48]$ and
  $\calcium[40]+\calcium[40]$ collisions.  The extracted isoscaling
  parameters are $\alpha=1.02\pm0.05$ and $\beta=-1.19\pm0.06$ for the
  Gogny force, and $\alpha=0.89\pm0.04$ and $\beta=-1.05\pm0.05$ for
  the Gogny-AS force.  }
\end{figure}

Since fragments are produced in a rapidly evolving system in the AMD
simulation, it is not evident \textit{a priori} whether the isoscaling
[Eq.\ (\ref{eq:isoscaling})] is expected in the fragment yield ratio.
Nevertheless, when we plot the fragment yield ratio,
$Y_2(N,Z)/Y_1(N,Z)$, between two reaction systems, we observe a clear
isoscaling relation, as shown in Figs.\ \ref{fig:isoscale2-6040} and
\ref{fig:isoscale2-4840} obtained for the fragments present at $t=300$
fm/$c$.  The extracted scaling parameters $\alpha$ and $\beta$ are
provided in the figure captions and indicated in individual panels.
The isoscaling parameters depend on the asymmetry term of the
effective force.  The fitting parameters, $\alpha$ and $\beta$, are of
larger magnitude when the Gogny force is used.  Moreover, $\alpha$
increases with increased differences in the asymmetry of the two
systems.

Given that neutron emission costs less energy in a more neutron-rich
system and proton emission costs more, it is not surprising that the
isospin fractionation is observed in the dynamical simulations.
However, the isoscaling is a nontrivial result, difficult to explain
outside of statistical considerations.  To explore the aspect of
equilibrium in fragment emission in AMD simulations, we further
explore the relation between the isoscaling and the fragment isospin
asymmetry in equilibrium and in simulations.



In the context of the expanding emitting source model \cite{EES}, it
has been pointed out \cite{TSANG-statmodel} that the isoscaling
parameter $\alpha$, obtained from the yield ratio of the emitted
fragments, is related to the $(Z_i/A_i)^2$ of an equilibrated emitting
source by
\begin{equation}
\alpha=4C_\text{sym}[(Z_1/A_1)^2-(Z_2/A_2)^2]/T,
\label{eq:linearrel-EES}
\end{equation}
where $C_\text{sym}$ is the symmetry energy and $T$ is the source
temperature.  However, in the AMD simulations of collisions there are
no easily identifiable emitting sources.  All fragments are emitted on
about equal footing.  To remedy the situation, we show that, for an
equilibrated system, Eq.\ (\ref{eq:linearrel-EES}) holds also when
$Z_i/A_i$ is replaced by $Z/\bar{A}_i$, where $\bar{A}_i$ is the
average mass number for fragment charge number $Z$ in system $i$,
provided that fragment properties change gradually with nucleon
content.

When we consider a system in equilibrium at the temperature $T$ and
pressure $P$, the number (or yield) of nucleus composed of $N$
neutrons and $Z$ protons is given by
\begin{equation}
Y_i(N,Z)=Y_{0i}\exp\bigl[-\bigl(G_\text{nuc}(N,Z)
                 -\mu_{ni}N-\mu_{pi}Z\bigr)/T\bigr],
\label{eq:YNZ}
\end{equation}
where the index $i$ specifies the reaction system, with the total
neutron and proton numbers $N^\text{tot}_i$ and $Z^\text{tot}_i$, and
$G_\text{nuc}(N,Z)$ stands for the internal Gibbs free energy of the
$(N,Z)$ nucleus.  The net Gibbs free energy $G_\text{tot}$ for the
system is related to the chemical potentials $\mu_{ni}$ and $\mu_{pi}$
by $G_\text{tot}= \mu_{ni}N^\text{tot}_i +\mu_{pi}Z^\text{tot}_i$.  In
Eq.\ (\ref{eq:YNZ}) and the following equations, we suppress the
$(T,P)$ dependence for different quantities including $G_\text{nuc}$,
$\mu_{ni}$ and $\mu_{pi}$.  It is clear that isoscaling [Eq.\
(\ref{eq:isoscaling})] is satisfied for Eq.\ (\ref{eq:YNZ}), with
$\alpha=(\mu_{n2}-\mu_{n1})/T$ and $\beta=(\mu_{p2}-\mu_{p1})/T$, as
long as the two systems have common temperature and pressure.

For each given $Z$, the dependence of $G_\text{nuc}$ on $N$, assuming
gradual changes, takes the form
\begin{equation}
G_\text{nuc}(N,Z)=a(Z)N+b(Z)+C(Z)(N-Z)^2/A.
\label{eq:GLD}
\end{equation}
Because the important range of $N$ is limited for a given $Z$, this
expansion is practically sufficient even when $G_\text{nuc}$ contains
surface terms, Coulomb terms, and any other terms which are smooth in
$A$ as \textit{e.g.} the term $\tau T \ln A$ introduced by Fisher
\cite{FISHER}.  We can regard $C(Z)$ as the symmetry energy
coefficient in the usual sense, because the second order terms in $N$
from the other terms are very small.  This fact can be proved by a
straightforward analytical calculation if a typical liquid-drop
mass-formula is assumed as an example.

Let us consider the average neutron number $\bar{N}_i(Z)$ of each
element $Z$.  By identifying the average value with the maximum of the
$N$ distribution [Eq.\ (\ref{eq:YNZ})], we get
\begin{equation}
(\partial/\partial N)[G_\text{nuc}(N,Z)
            -\mu_{ni}N-\mu_{pi}Z]|_{N=\bar{N}_i(Z)}=0.
\end{equation}
A straightforward calculation, using the specific form of
$G_\text{nuc}$ of Eq.\ (\ref{eq:GLD}), results in
\begin{equation}
C(Z)\bigl[1-4\bigl(Z/\bar{A}_i(Z)\bigr)^2\bigr]=\mu_{ni}-a(Z),
\end{equation}
with $\bar{A}_i(Z)=Z+\bar{N}_i(Z)$.  The equations for the two
reaction systems, $i=1$ and $i=2$, subtracted side by side then yield
\begin{equation}
\frac{\alpha}
     {\bigl(Z/\bar{A}_1(Z)\bigr)^2-\bigl(Z/\bar{A}_2(Z)\bigr)^2}
=4C(Z)/T,
\label{eq:linearrel}
\end{equation}
relating the isoscaling parameter $\alpha$, the $(Z/A)^2$ of fragments
and the symmetry energy coefficient $C(Z)$ which is a function of
$(T,P)$.  An interesting fact is that this relation does not depend on
the terms in $G_\text{nuc}$ other than the symmetry energy term.


\begin{figure}
\includegraphics[width=0.48\textwidth]{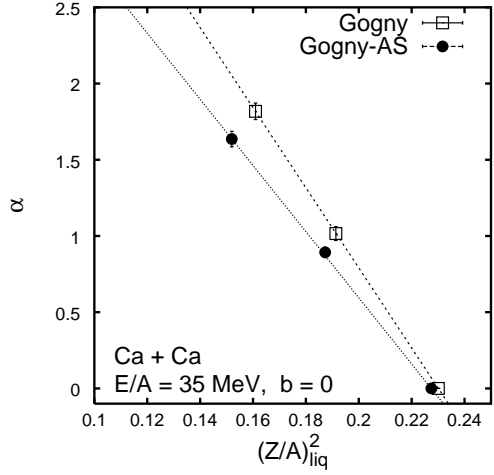}
\caption{\label{fig:residue-alpha} Relation between
$(Z/A)^2_\text{liq}$ and $\alpha$ for the three systems
$\calcium[60]+\calcium[60]$, $\calcium[48]+\calcium[48]$ and
$\calcium[60]+\calcium[60]$ (from left to right).  Open squares and
filled circles show the results of AMD with Gogny force and Gogny-AS
force, respectively.  Both $(Z/A)^2_\text{liq}$ and $\alpha$ are
calculated for fragments recognized at $t=300$ fm/$c$.  The straight
lines are drawn so as to connect the points for
$\calcium[40]+\calcium[40]$ and $\calcium[60]+\calcium[60]$. Their
slopes are $-26.27$ for Gogny force and $-21.70$ for Gogny-AS force.}
\end{figure}

Both sides of Eq.\ (\ref{eq:linearrel}) depend on the fragment charge
$Z$, in principle, and can provide information of the surface effect
in the symmetry energy.  This issue will be pursued in a separate
paper.  In the present letter, we use the fact the $Z$-dependence of
$Z/\bar{A}_i(Z)$ in simulations is weak for $Z\gtrsim5$, making it
meaningful to consider the isospin asymmetry of the liquid part
$(Z/A)^2_\text{liq}\equiv (Z_\text{liq}/A_\text{liq})^2$ that has been
averaged over all the fragments with $A>4$.  For this quantity we
expect the relation
\begin{equation}
\frac{\alpha}
   {\bigl(Z/A\bigr)^2_{\text{liq},1}-\bigl(Z/A\bigr)^2_{\text{liq},2}}
=4C/T.
\label{eq:linearrel-liq}
\end{equation}

Let us check whether this equilibrium relation
(\ref{eq:linearrel-liq}) is satisfied by the AMD simulations that do
not incorporate any assumption of equilibrium.  Figure
\ref{fig:residue-alpha} shows the correlation in the AMD simulations
between $(Z/A)^2_\text{liq}$ and the isoscaling parameter $\alpha$
from Figs.\ \ref{fig:isoscale2-6040} and \ref{fig:isoscale2-4840} for
the three reaction systems.  A linear relation is observed in
accordance to the equilibrium relation (\ref{eq:linearrel-liq}),
suggesting applicability of statistical laws to the isospin
composition of fragments.  The extracted value of $4C/T$ depends on
the time of the fragment recognition, but the dependence is weak as
both $\alpha$ and $(Z/A)^2_{\text{liq},1}-(Z/A)^2_{\text{liq},2}$ are
decreasing functions of $t$ for $t\gtrsim200$ fm/$c$. From the slope
of the linear relation, we obtain $4C/T = 26.3$ for Gogny force and
$4C/T=21.7$ for Gogny-AS force.  The ratio of
$C(\text{Gogny})/C(\text{Gogny-AS})\approx 5/4$, from Fig.\
\ref{fig:symeng}, is consistent with the idea that fragmentation
occurs at low density, $\rho<\rho_0$.  The absolute values of $C$ of
16-20 MeV and 22-26 MeV are reasonable assuming the temperatures
$T\sim3$-4 MeV.

In conclusion, a linear relation is expected between the isoscaling
parameter and the fragment isospin asymmetry $(Z/A)^2$ under
statistical assumptions.  Isoscaling is observed in the dynamical AMD
simulation and the results well comply with the linear relation,
suggesting that the fragment isospin composition is subject to the
statistical laws even in a dynamical picture of production.  The slope
of this linear function yields information on the symmetry energy
for the fragments.

\begin{acknowledgments}
This work was supported by Japan Society for the Promotions of Science
and the US National Science Foundation under the U.S.-Japan
Cooperative Science Program (INT-0124186), by the High Energy
Accelerator Research Organization (KEK) as the Supercomputer Project
No.~83 (FY2002), and by grants from the US National Science
Foundation, PHY-0070818, PHY-0070161 and PHY-01-10253.  The work was
also partially supported by RIKEN as a nuclear theory project.
\end{acknowledgments}

\end{document}